\documentclass{PoS}

\title{Topological charge membranes and Goldstone boson propagation in QCD}

\ShortTitle{Goldstone boson propagation}

\author{\speaker{H. B. Thacker}\thanks{A footnote may follow.}\\
        Physics Dept., University of Virginia, Charlottesville VA\\
        E-mail: \email{hbt8r@virginia.edu}}

\abstract{Both theoretical arguments and Monte Carlo observations indicate that the topological structure of the QCD
vacuum consists of a laminated array of extended, coherent codimension-one membranes of alternating sign. Large-$N_c$
arguments, supported by gauge/string holography, indicate that these membranes are domain walls which separate discrete
``flux vacua'' with values of the topological $\theta$ parameter which differ by $\pm 2\pi$. This exposes a close analogy
with 2D $U(1)$ gauge theory, where $\theta$ can be interpreted as electric polarization, and the domain walls are 
pointlike charged particles. In 4D QCD, the $\theta$ parameter represents background Ramond-Ramond flux, which can be interpreted
as a polarization of the charged membranes in the vacuum. In this framework, the chiral condensate is formed from the quark surface
modes on the membranes. Massless Goldstone boson propagation takes place due to a coordination between bulk oscillations
of the polarization field $\theta$ and the surface currents represented by the Chern-Simons 3-form on the brane surface. 
This coordination is enforced by overall gauge invariance which imposes an anomaly inflow constraint between bulk and surface currents.}

\FullConference{The 7th International Workshop on Chiral Dynamics,\\
		August 6 -10, 2012\\
		Jefferson Lab, Newport News, Virginia, USA}

\begin{document}
\section {Introduction}
The role of gauge field topology in hadron phenomenology appears most directly in the resolution of the ``U(1) problem,'' where topologically nontrivial
gauge configurations give rise to the axial anomaly and are responsible for giving a large mass to the flavor singlet pseudoscalar $\eta'$ meson. But it
is also likely that the topological structure of the QCD vacuum plays a central role in spontaneous chiral symmetry breaking and Goldstone boson propagation.
The idea that the formation of the chiral condensate in QCD is driven by topological fluctuations of the gauge field has been extensively explored.
The most detailed theoretical studies of this issue have been carried out in the context of the instanton liquid
model \cite{Diakonov84,Shuryak}. In that model, the localized 'tHooft zero modes of the instantons mix and form a band of low-lying Dirac eigenmodes which are occupied by the
chiral condensate. However, it was pointed out long ago by Witten \cite{Witten79} that the $U(1)$ problem did not require instantons, but could be resolved in any model that incorporated
nonzero topological susceptibility in the pure glue theory. Not only are instantons an unnecessary restriction on the form of topological
fluctuations, but the behavior of the $\eta'$ mass (as a function of $N_c$) 
predicted by the instanton model conflicts with chiral phenomenology in the large-$N_c$ limit. The large-$N_c$ arguments also lead to a qualitatively different
picture of the QCD vacuum, consisting of discrete quasivacua separated by extended 2+1 dimensional topological membranes (domain walls) rather than a gas or liquid of localized instantons. The fact that membranes fitting
this description have been observed in Monte Carlo studies of pure glue $SU(3)$ gauge theory \cite{Horvath03,Ilgenfritz} lends strong support to the large-$N_c$
view of the QCD vacuum.

The connection between topological charge fluctuations and the formation of the chiral condensate is undoubtedly more general than the instanton liquid model.
In the instanton picture, the 'tHooft zero modes are localized lumps of left and right chiral charge $\bar{q}(1\pm\gamma_5)q$,
around instantons and antiinstantons, respectively. But we expect that generic positive or negative fluctuations of topological charge will produce
low lying left- and right-handed Dirac states attached to the fluctuations, providing the eigenmodes required to form a condensate. 
The membranes that appear in QCD Monte Carlo configurations are 
dipole layers of topological charge, with positive and negative charge sheets on opposite sides of the membrane. This configuration of
topological charge around the brane gives rise to left- and right-handed quark surface modes on opposite sides of the brane. In a vacuum with a finite density of branes, 
these quark surface modes form the chiral condensate. Moreover, the extended, coherent structure of the topological charge sheets allows for delocalized (i.e. propagating) quark surface modes. These
delocalized modes provide a natural mechanism for Goldstone boson propagation.

\section {Topological membranes in gauge theory}
It is instructive to first consider the role of gauge topology and domain walls in a 2-dimensional theory with a $U(1)$ gauge invariance, 
specifically the $CP^{N-1}$ sigma model. This model exhibits many of the same 
topological properties as QCD. From lattice Monte Carlo studies\cite{Ahmad05,Lian07,Keith-Hynes08} it has been shown that, while the vacuum of $CP^1$ and $CP^2$ is dominated
by small localized instantons, for $N>3$ the topological fluctuations are predominantly in the form of extended codimension-one membranes, 
quite analogous to those seen in 4D $SU(3)$ gauge theory. 

The membranes that appear in large-N QCD \cite{Witten79} are
domain walls which separate discrete, quasi-stable vacua having topological $\theta$ parameters which differ by $\pm2\pi$. 
We can introduce such a membrane by including in the path integral a 
space-dependent $\theta$ term where $\theta$ is a step function on the membrane surface. 
\begin{equation}
{\cal L}_{YM}\rightarrow {\cal L}_{YM} + \int\theta(x)Q(x)d^Dx
\end{equation}
with $\theta(x)=\theta_0$ on one side of the brane and 0 on the other side.
For both 2D U(1) theories and 4D Yang-Mills, the topological charge density $Q(x)$ can be written as the divergence of a Chern-Simons current, $Q(x)\propto \partial^{\mu}K_{\mu}$. 
Thus, with an integration by parts, the action of the brane reduces to an integral of the Chern-Simons $(D-1)$ form over the surface of the membrane. In the 2D U(1) example, the CS
current is just the dual of the gauge potential $K_{\mu}=\varepsilon_{\mu\nu}A^{\nu}$. In this case, the action of the membrane is just the ordinary Wilson line integral,
\begin{equation}
\label{eq:2dbrane}
{\cal L}_{brane} = \frac{\theta_0}{2\pi}\int A_{\mu}dx^{\mu}
\end{equation}
So in the 2D U(1) theory, the topological membrane can be interpreted as the world line of a point charge. This incorporates Coleman's original interpretation
of the $\theta$ parameter in the massive Schwinger model (QED2) as a background electric field. In one space dimension, a charged particle plays the role of a domain wall 
separating vacua which differ by one unit of electric flux. For 2D $U(1)$ theories, the analog of the discrete vacua in large $N_c$ QCD are ``flux vacua'' with a given number 
of units of background electric flux. The $\pm 2\pi$ quantization of the step in $\theta$ across a domain wall is simply Gauss's law combined with the quantization of electric charge.

For 4-dimensional QCD, the action of a brane is given by the ``Wilson bag'' integral of the Chern-Simons tensor over the 3-dimensional world volume of the brane \cite{Luscher78},
\begin{equation}
\label{eq:CS_action}
{\cal L}_{brane}=\frac{\theta_0}{2\pi}\int K_3^{\mu\nu\lambda}dx_{\mu}\wedge dx_{\nu}\wedge dx_{\lambda}
\end{equation}
where $K_3$ is the Chern-Simons 3-form,
\begin{equation}
\label{eq:CScurrent}
K_{\mu} = \varepsilon_{\mu\alpha\beta\gamma}{\rm Tr}\left(A^{\alpha}\partial^{\beta}A^{\gamma} + \frac{2}{3}A^{\alpha}A^{\beta}A^{\gamma}\right)
\equiv \varepsilon_{\mu\alpha\beta\gamma}K_3^{\alpha\beta\gamma}
\end{equation}
Just as the Wilson line represents a charged point particle, (\ref{eq:CS_action}) represents an membrane-like excitation of the CS tensor 
which depends only on the gauge field components within the 3D world volume of the brane. 
Like the Wilson line,
the CS action (\ref{eq:CS_action}) is not locally gauge invariant, and the action integrated over a closed surface is only invariant mod $2\pi$ 
under a topologically nontrivial gauge transformation on the brane surface. In 2D $U(1)$, $\delta A_{\mu}=\partial_{\mu}\omega$, and the mod $2\pi$ ambiguity of
the action corresponds to the winding number of the gauge phase $e^{i\omega}$ around the loop. Similarly, in 4D Yang-Mills, the gauge variation of the CS 3-form $K_3$ is the exterior derivative
of a WZW-like 2-form $K_2$ on a fixed-time 2-dimensional brane surface,
\begin{equation}
\delta K_{\mu} = \varepsilon_{\mu\nu\alpha\beta}\partial^{\nu}K_2^{\alpha\beta} \;.
\end{equation}
This 2-form $K_2$ is the 4D Yang-Mills analog of the gauge phase $\omega$ along the Wilson line in (\ref{eq:2dbrane}).
It plays the role of a Hamiltonian describing the time-dependent 2-dimensional brane surface. The WZW 2-form $K_2$ is a 
functional of both the Yang-Mills field and the gauge transformation $g=e^{i\omega}$ used 
to define $\delta K_3$. As discussed in \cite{TX12}, small nontopological gauge variations on the brane surface are related by anomaly inflow to transverse fluctuations
of the membrane surface. This defines an interplay between bulk space-time fluctuations of the brane surfaces and the fluctuation of surface color currents on the brane. This connection
is central to the dynamics of Goldstone boson propagation.

In the $CP^{N-1}$ sigma model, the $U(1)$ gauge potential $A_{\mu}$ is an auxiliary field with no kinetic term at tree level. Its EOM sets it equal to the $U(1)$ current $j_{\mu}$.
In the large $N$ approximation, an $F_{\mu\nu}^2$ term is generated in the action by one-loop vacuum polarization. As a bulk field, the
value of $\theta(x)$ in the 2-dimensional $CP^{N-1}$ model can thus be interpreted as the local electric polarization of charged pairs in the
vacuum. As in a dielectric medium, the local electric charge density is given by the spatial variation of the polarization, while a time-varying polarization produces a polarization current:
\begin{equation}
\label{eq:bosonize}
\partial_1\theta = j_0 \;,\;\; \partial_0\theta = -j_1
\end{equation}
In a 2D U(1) theory in which electric charge is carried by an elementary fermion field (QED2), $j_{\mu}=\bar{\psi}\gamma_{\mu}\psi$. So in this case, the
polarization field $\theta(x)$ is just the usual bosonization of the current in 2D,
\begin{equation}
\partial_{\mu}\theta=\epsilon_{\mu\nu}j_{\mu}
\end{equation}
For a step function $\theta(x)$ at a domain wall, this specifies that the discontinuity of $\theta$ is associated with an edge current along the domain boundary. 
Noting that the axial-vector current is just the dual of the vector current, 
\begin{equation}
\label{eq:dual}
j_{\mu}=\varepsilon_{\mu\nu}j_5^{\nu}
\end{equation}
we see that the polarization field $\theta(x)$ is a chiral Goldstone field,
\begin{equation}
\label{eq:goldstone}
\partial^{\mu}\theta = j_5^{\mu}
\end{equation}

This relation between domain walls and the $U(1)$ Goldstone field generalizes to the case of QCD in 4 dimensions. While in the 2-dimensional U(1) theory $\theta$ is a polarization field
describing charged particle pairs, in 4-dimensional QCD $\theta$ describes the polarization of membrane-antimembrane pairs.
From both theoretical arguments \cite{Thacker10} and Monte Carlo observations \cite{Horvath05,Ilgenfritz} we expect the local structure of the QCD
vacuum to look like a ``tachyonic crystal,'' a laminated array of alternating--sign membranes which are locally flat and parallel over distances small compared to the confinement scale. 
(The confinement scale is determined by the distance over which the orientation of the branes decorrelates.) Just as electric polarization in the $CP^{N-1}$ vacuum arises
from the presence of polarizable charged particle pairs, similarly topological susceptibility in QCD arises from the presence of polarizable membranes in the vacuum.
In addition to providing finite $\chi_t$ and resolving the $U(1)$ problem, this condensate of membranes points to a fundamental understanding of Goldstone boson propagation.
In a theory with one massless quark flavor, a rotation of $\theta$ is equivalent to a $U(1)$ chiral rotation. This identifies the brane polarization field $\theta(x)$ as
the chiral field associated with the $\eta'$ meson.
In the large $N_c$ limit of QCD, the $\eta'$ becomes an approximate Goldstone boson, with a mass of order $1/N_c$. So in this limit we expect the $\eta'$ to propagate
via the same quark eigenmodes that allow propagation of massless pions. As I discuss in the next Section, the propagation of Goldstone bosons in the QCD vacuum 
is a result of an interplay between the transverse motion of the branes, described by the bulk polarization field $\theta(x)$, and the surface color currents associated
with the Chern-Simons action on the brane. This interplay is best understood in terms of the descent equations of Yang-Mills theory \cite{Faddeev84} which define the
interrelationship between gauge variations and exterior derivatives of operators descended from the topological charge $Q(x)\propto Tr[F\tilde{F}]$. The key physical
idea here is that of ``anomaly inflow'' \cite{Callan-Harvey}, which relates bulk and boundary currents at a membrane surface. 

\section{Goldstone boson propagation and anomaly inflow}

The ``secret long-range order'' in QCD, first pointed out by Luscher \cite{Luscher78}, is seen from the observation that nonzero topological susceptibility {\it requires}
the existence of a massless pole in the Chern-Simons currrent correlator,
\begin{equation}
\label{eq:RRprop}
\int d^4x e^{iqx}\langle K_{\mu}(x) K_{\nu}(0)\rangle \stackrel{q\rightarrow 0}{\sim} \frac{q_{\mu}q_{\nu}}{(q^2)^2}\chi_t + \ldots
\end{equation}
where the other terms are ones that vanish when dotted with $q^{\mu}$ or $q^{\nu}$. 
For $N_f\leq 1$ QCD has a finite mass gap, so the pole in (\ref{eq:RRprop}) must not appear in any physical amplitude. There is a plausible escape, since
$K_{\mu}$ itself is not a gauge invariant operator. On the other hand, the residue of the pole in (\ref{eq:RRprop}) is the gauge invariant physical quantity $\chi_t$, so the massless
pole cannot be simply gauge transformed away. To see how this problem resolves itself, it is again instructive to look at a 2D U(1) example, namely QED2 \cite{Kogut75}.
In the Lorentz gauge formulation of this model, the pole in the CS correlator (\ref{eq:RRprop}) is cancelled (screened) in physical amplitudes by a massless ghost pole which represents
a quark-antiquark Goldstone mode. The QED2 cancellation between the Goldstone pole and the gauge pole in the CS correlator (\ref{eq:RRprop}) reflects the separation
of the gauge invariant axial vector current $\hat{j}_5^{\mu}$ into a quark-antiquark Goldstone boson term $j_5^{\mu}=\partial_{\mu}\theta$ which is exactly conserved 
and a Chern-Simons term which accounts for the anomaly,
\begin{equation}
\label{eq:inflow}
\hat{j}_5^{\mu} = \partial^{\mu}\theta + K^{\mu}
\end{equation}
This same separation occurs in QCD, with the Chern-Simons current given by (\ref{eq:CScurrent}).
The massless poles coupling to the two operators on the right cancel in matrix elements of the gauge invariant current. At the surface of a CS brane, the requirement
that the right-hand side of (\ref{eq:inflow}) is gauge invariant may
be viewed as a bulk-boundary or anomaly inflow constraint relating the gauge variation of the polarization field $\theta$ across the brane to that of the CS current on the brane surface.
It specifies that $\theta(x)$ must transform nontrivially under color gauge transformations
\begin{equation}
\label{eq:gi}
\delta(\partial^{\mu}\theta) = -\delta K_{\mu} = \varepsilon_{\mu\nu\alpha\beta}\partial^{\nu}K_2^{\alpha\beta}
\end{equation}
In practical calculations, this pole cancellation is reflected in the cancellation between the quark-antiquark (valence) and gluonic (hairpin) diagrams in the physical $\eta'$ correlator, 
These diagrams separately have massless poles in the chiral limit, and the cancellation of their long-range components is a well-known and striking
numerical phenomenon in the lattice QCD calculation of the flavor singlet pseudoscalar correlator. We can now introduce two massless quark flavors and construct the
nonsinglet pion correlator, obtained by simply discarding the hairpin ($q\bar{q}$ annihilation) diagram. This shows that a true Goldstone pion is a propagating polarization wave.
The cancellation of massless poles in the $\eta'$ correlator identifies the nonsinglet axial current on the brane surface with the Chern-Simons current. Note that, even in the
absence of $q\bar{q}$ annihilation, the
nonsinglet axial current consists of both bulk and boundary terms. (The $\theta$ field is discontinuous at the brane surface, whether or not $q\bar{q}$ annihilation
is allowed.) The gauge invariance or anomaly inflow constraint (\ref{eq:gi}) at the brane surfaces enforces the overall conservation of bulk + boundary currents and hence the masslessness of
nonsinglet Goldstone boson propagation.

\end{document}